\documentstyle[preprint,aps]{revtex}
\begin{document}
\title{
Generic Emergence of Power Law Distributions and L\'evy-Stable
Intermittent Fluctuations in Discrete Logistic Systems
}
\author{
Ofer Biham,
Ofer Malcai,
Moshe Levy
and
Sorin Solomon
}
\address{
Racah Institute of Physics, The Hebrew University, Jerusalem
91904, Israel \\
{\small
biham@flounder.fiz.huji.ac.il, malcai@flounder.fiz.huji.ac.il, shiki@cc.huji.ac.il,
sorin@vms.huji.ac.il}}
\maketitle

\begin{abstract}
The dynamics of
generic stochastic
Lotka-Volterra (discrete logistic) systems of the form \cite{Solomon96a}
$w_i (t+1) = \lambda(t) w_i (t) + a {\bar w (t)} - b w_i (t) {\bar w(t)}$
is studied by computer simulations.
The variables
$w_i$, $i=1,...N$,
are the individual system components and
${ \bar w (t)} = {1\over N}  \sum_i  w_i (t)$
is their average.
The parameters $a$ and $b$ are constants,
while $\lambda(t)$ is randomly chosen at
each time step from a given distribution.
Models of this type describe the temporal evolution of a large variety
of systems such as stock markets and city populations.
These systems are characterized by a large number of interacting objects
and the dynamics is dominated by multiplicative processes.
The
instantaneous probability
distribution $P(w,t)$
of the system components $w_i$,
turns out to fulfill a (truncated) Pareto power-law
$P(w,t) \sim w^{-1-\alpha}$.
The time evolution of
${ \bar w (t)} $
presents intermittent fluctuations
parametrized by a truncated L\'evy distribution
of index $\alpha$,
showing a connection between the
distribution of the $w_i$'s at a given time and the
temporal fluctuations of their average.
\end{abstract}


\newpage

\section{Introduction}

Power-law distributions have been observed
in all domains of the
natural sciences
as well as in economics,
linguistics
and many
other fields.
Widely studied examples of power law distributions include
the energy distribution between scales in
turbulence
\cite{Kolmogorov62},
distribution of earthquake magnitudes
\cite{Gutenberg56},
diameter distribution of craters and asteroids
\cite{Mizutani80},
the distribution of city populations
\cite{Zipf,Zanette97},
the distributions of income
and of wealth
\cite{Pareto,Mandelbrot61,Mandelbrot51,Mandelbrot63,Atkinson78},
the size-distribution of business firms
\cite{Simon58,Simon77}
and the distribution of the frequency of appearance
of words in texts
\cite{Zipf}.
A related phenomenon is the fact that in a variety of systems the
temporal fluctuations exhibit a scale invariant behavior in the
form of (truncated) L\'evy-stable distributions.
Well known examples are the fluctuations in stock markets
\cite{Mandelbrot61,Bak97}.

Although systems
which exhibit power-law distributions
have been studied extensively in recent years
there is no universally accepted
framework which can explain
the origin of the abundance and diversity of power-law distributions.
One context in which the emergence of scaling laws
and long range correlations in space and time
is well understood
is equilibrium statistical physics at the critical point
\cite{Kadanoff66,Fisher74,Wilson75,Gell-Mann}.
By contrast,
scaling behavior, power law distributions
as well as spatial and temporal power law correlations
in {\bf generic}
natural systems is still the subject of intense study
\cite{Mandelbrot,Cizeau143,Cizeau342,Mantegna94,Mantegna97,Mantegna96,Mantegna95,Stanley95,Stanley96,Amaral97a,Amaral97b,Bak87,Bak89b,Bak90}.

An approach that proved to be useful in the study of complex systems
is to identify for each system the relevant elementary
degrees of freedom and their interactions and to follow-up
(by monitoring their computer simulation)
the emergence in the system of the
macroscopic collective phenomena
\cite{Solomon95b}.
This approach was applied to the study of
multiscale dynamics in spin glasses
\cite{Persky96}
and
stock market dynamics
\cite{Levy96b}.
Using a generic class of models with a
large number of interacting degrees of
freedom, it was shown that
macroscopic dynamics emerges
under rather general conditions.
This dynamics exhibits power law scaling
as well as intermittency
\cite{Levy96b,Levy96a,Solomon96a}.
These models turn out to be particularly suitable to describe systems such
as stock market dynamics with many individual investors
\cite{Stauffer98,Hellthale95,Anderson88,Zhang118,Challet006,marsili97}
where each system component describes a single investor (or stock \cite{maslov98}).
Such systems involve complex temporal dynamics of
many degrees of freedom but no spatial structure.
The models introduced in  
\cite{Solomon96a} can also describe systems such as
population dynamics
\cite{May76,Hubermann97,Weisbuch97,Pimm91,marsili98,Zanette97},
spatial domains
in magnetic
\cite{Yuhai002}
or
turbulence
models
\cite{Peinke77,Peinke97}
or regions in generic phase spaces
\cite{Asselmeyer003,Ebeling002,Ebeling005},
which have spatial dependence.

In this paper we present numerical studies of
generic stochastic Lotka-Volterra systems.
These systems \cite{Solomon96a} basically consist of coupled
dynamic equations which describe the discrete time evolution
of the basic system components
$w_i$, $i=1,\dots,N$.
The structure of these equations resembles the
logistic map and they are coupled through the average value
${ \bar w (t)} $.
The dynamics includes autocatalysis both at the individual level
and at the community level as well as a saturation term.
We find that under very general conditions, the
system components spontaneously evolve into a
power-law distribution
$P(w,t) \sim w^{-1-\alpha}$.
The time evolution of
${ \bar w (t)} $
presents intermittent fluctuations
parametrized by a (truncated) L\'evy-stable distribution
with the same index $\alpha$,
showing an intricate relation between the
instantaneous
distribution
of the system components
and the
temporal fluctuations
of their average.

The paper is organized as follows.
In Sec. II we present the generalized logistic model introduced in 
\cite{Solomon96a}.
Simulations and results are reported in Sec. III.
Discussion of previous results as well as of our
findings is given in Sec. IV, and a summary in Sec. V.

\section{The Model}

The generalized logistic system \cite{Solomon96a}
describes the evolution in discrete time of
$N$ dynamic variables
$w_i$, $i=1,\dots,N$.
At each time step $t$,
an integer $i$ is chosen randomly in the range
$1 \leq i \leq N$, which is the index of the dynamic variable $w_i$
to be updated at that time step.
A random multiplicative factor
$\lambda(t)$ is then drawn from a given distribution
$\Pi(\lambda)$, which is independent of $i$ and $t$.
This can be, for example, a uniform distribution in the range
$\lambda_{min} \leq \lambda \leq \lambda_{max}$,
where
$\lambda_{min}$
and
$\lambda_{max}$
are predefined limits.
The system is then updated according to

\begin{eqnarray}
w_i (t+1) &=&   \lambda (t)  w_i (t) +   a \bar w (t) - b  w_i (t) \bar w(t)  \nonumber \\
w_j (t+1) &=& w_j (t), \ \ \ \ \ \ j=1,\dots,N; \ j \ne i.
\label{gedilo}
\end{eqnarray}

\noindent
This is an asynchronous update mechanism.
The average
value of the system components
at time t, is given by

\begin{equation}
\bar w(t) = {1 \over N} \sum_{i=1}^N w_i(t).
\end{equation}

\noindent
In general, using instead of the average,
a weighted average
of the
$w_i$'s
would lead to similar results.
The parameters $a$ and $b$ may, in general, be slowly varying functions
of time, however we will now consider them as constants.
The first term on the right hand side of
Eq. (\ref{gedilo}),
describes the effect of auto-catalysis at the individual level.
For instance, in a stock-market system
it represents the increase (or decrease) by a random factor
$\lambda(t)$
of the capital of the investor
$i$ between time $t$ and $t+1$.
The second term in
Eq. (\ref{gedilo}),
describes the effect of  auto-catalysis at the community level.
In an economic model, this term can be related to the social
security policy or to general publicly funded services which every individual receives.
In molecular or magnetic systems, this term may represent the mean-field
approximation to the effect of diffusion or convection
\cite{Yuhai002}.
The third term in
Eq. (\ref{gedilo}),
describes  saturation or the competition for limited resources.
In an ecological model, this term implies that for large enough
densities, the population starts to exhaust the available resources and each
sub-population loses from the competition over resources a term proportional to
the product between the average density population and its own size.
We refer to
Eq. (\ref{gedilo})
as the generalized discrete logistic (GL) system
because when averaged over $i$, this system gives the well known discrete logistic
(Lotka-Volterra)
equation
\cite{Lotka25,Volterra26}

\begin{equation}
w (t+1) = ( {\bar  \lambda } +a ) w (t) - b w^2 (t).
\end{equation}

\noindent
In the general case, the parameters
$a$, $b$ and  the distribution
$\Pi (\lambda)$
may depend on time.
Consequently, even the solution of the asymptotic stationarity condition
$\bar w (t+1) = \bar w (t)$
may depend on time according to
\begin{equation}
{\bar w (t)} = ({\bar \lambda }(t) + a - 1)/b(t).
\label{equil}
\end{equation}
In fact, the typical dynamics of microscopic market models
\cite{Levy94,Levy95a,Solomon96b,Levy97}
is generically {\it not} in a steady state.
As will be shown below,
systems which exhibit an effective GL dynamics
[Eq. (1)]
lead \cite{Solomon96a}, under very general conditions,
to a power law distribution of the values $w_i$:

\begin{equation}
P(w) \sim w^{-1-\alpha}.
\label{pareto}
\end{equation}

\noindent
Moreover, the time evolution of
${ \bar w (t)} $
presents intermittent fluctuations
following a (truncated) L\'evy-stable distribution
with the same index $\alpha$,

\section{Simulations and Results}

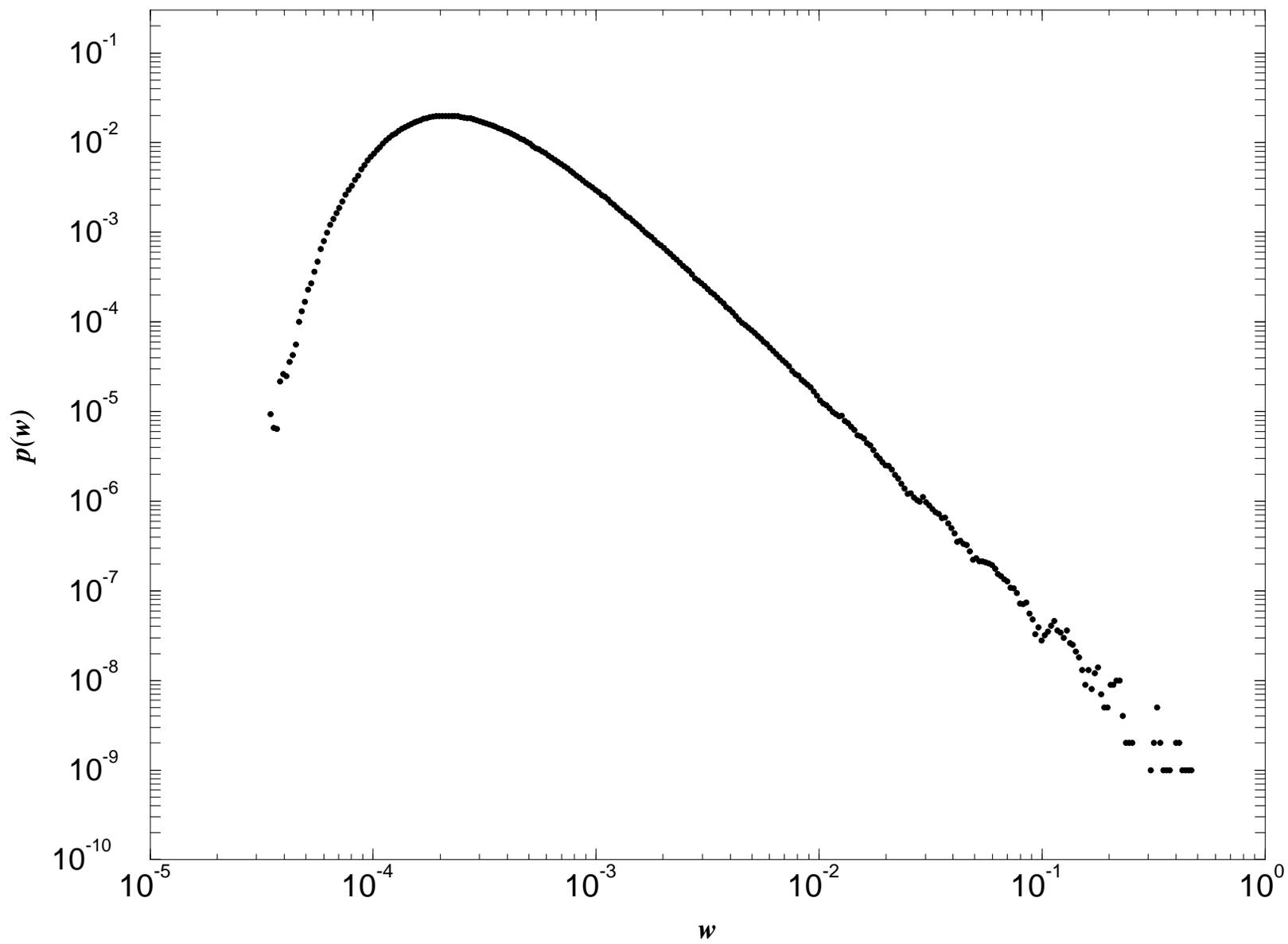
\begin{figure}
\caption{
The distribution of wealth $w_i$, $i=1,\dots,N$
(the number of investors $P(w)$ possessing wealth $w$)
for $N=1000$ investors obtained
from a numerical integration of Eq.(\ref{gedilo}) with parameters $a=0.00023$,
$b=0.01$ and $ \Pi(\lambda)$ uniformly distributed in the range $1.0 < \lambda < 1.1$.
The distribution
(presented here on a $\log - \log$ scale)
exhibits a knee on the left hand side and a broad tail of power law
distribution on the right hand side. This power law behavior is described
by $P(w) \sim w^{-1-\alpha}$, where the exponent $\alpha = 1.4$.
The distribution is bounded by an upper cutoff around $w_{max} = N \bar w$.
}
\label{fig:pareto}
\end{figure}

To examine the behavior of the GL model introduced in \cite{Solomon96a},  Eq.
(\ref{gedilo})
we performed extensive computer simulations.
Most simulations were done with $N=1000$
system components, using various values of the
parameters $a$ and $b$
and different distributions
$\Pi(\lambda)$
of the multiplicative
factor $\lambda$.
We focused on the power law distribution of the system components
$w_i$
as well as on the fluctuations of
$\bar w$.
Fig.
\ref{fig:pareto}
shows the distribution of
$w_i$, $i=1,\dots,N$,
obtained for
$N=1000$,
$a=0.00023$,
$b= 0.01$
and
$\lambda$
uniformly distributed in the range
$1.0 \leq \lambda \leq 1.1$.
A power law distribution is found within the range

\begin{equation}
\bar w < w_i  <  N \bar w
\label{trunk}
\end{equation}

\noindent
which is bounded from below by the average wealth and from above
by the total wealth,
and spans nearly three decades.

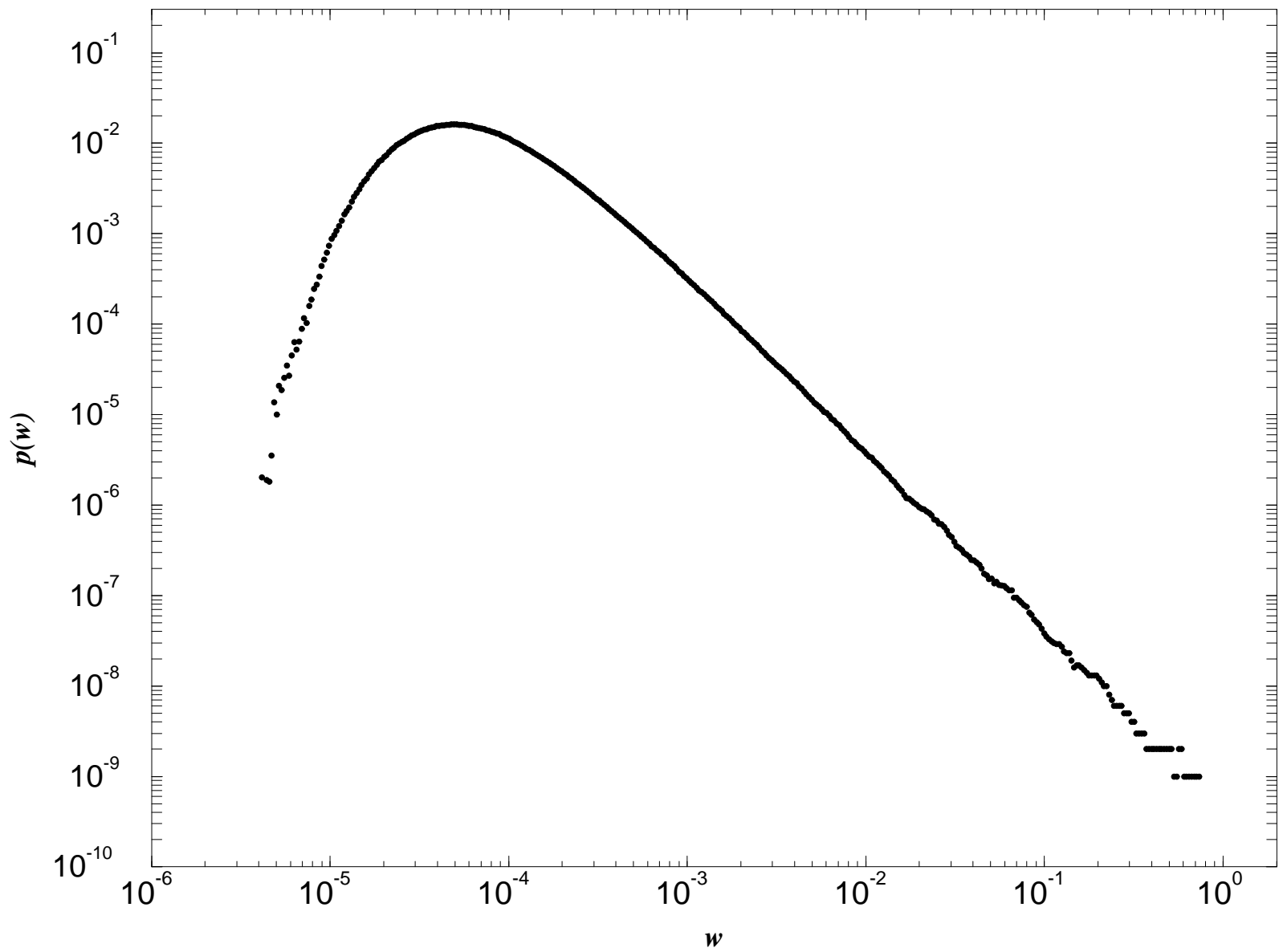
\begin{figure}
\caption{
The distribution of the values of $w_i$, $i=1,\dots,N$ for $N=1000$, $a= 0.0001$,
$b=0.0$ and $ \Pi(\lambda)$ uniformly distributed in the range $1.0 < \lambda < 1.1$.
Because of the absence of the saturation term ($b=0$),
the system is
not stationary and $\bar w$ varies in time by orders of magnitude.
In spite of this,
the instantaneous normalized $w$ distribution at each instant
remains always a power law of constant exponent $\alpha$.
}
\label{fig:bzero}
\end{figure}

The robust nature of the power-law distribution
is demonstrated
in
Fig.
\ref{fig:bzero}
for $b=0$.
In this case
$\bar w(t)$
does not reach a steady state and keeps increasing
(or decreasing) indefinitely.
However, the power-law behavior is maintained.
Moreover
we find that
the the exponent
$\alpha$
is insensitive to variations in $b$:
even for values of $b$ differing by an order of magnitude
(corresponding to $\bar w$
varying by an order of magnitude),
the power law exponent $\alpha$ is virtually unchanged.

At each time step the system component to be updated is chosen randomly.
Since the system components
$w_i$
exhibit a power-law distribution,
given by Eq.
(\ref{pareto}),
the impact of the update move on
$\bar w(t)$
exhibits a broad distribution.
The dynamics involves,
according Eq.
(\ref{gedilo}),
a generalized  random
walk with steps sizes distributed  according to
Eq.
(\ref{pareto}).
Therefore, the
stochastic fluctuations

\begin{equation}
r(\tau) = { {\bar w(t+\tau) - \bar w(t)} \over {\bar w(t)} }
\label{return}
\end{equation}

\noindent
of $\bar w (t)$
after
$\tau$
time steps, are governed by a truncated L\'evy stable
distribution
$L_{\alpha} (r)$.
This means that rather than shrinking like
$N^{\small {-1/2}}$
the fluctuations of
$\bar w (t) $
have infinite variance in the
thermodynamic limit
(modulo the truncation).
The truncation in the L\'evy-stable distribution corresponds to
the cutoffs in the power-law distribution,
given in Eq.
(\ref{trunk}).
Typically, the truncation in $r$
is
bounded by
the relative width of
$\lambda$
times the largest
$w_i/{(N \bar w)}$
value.

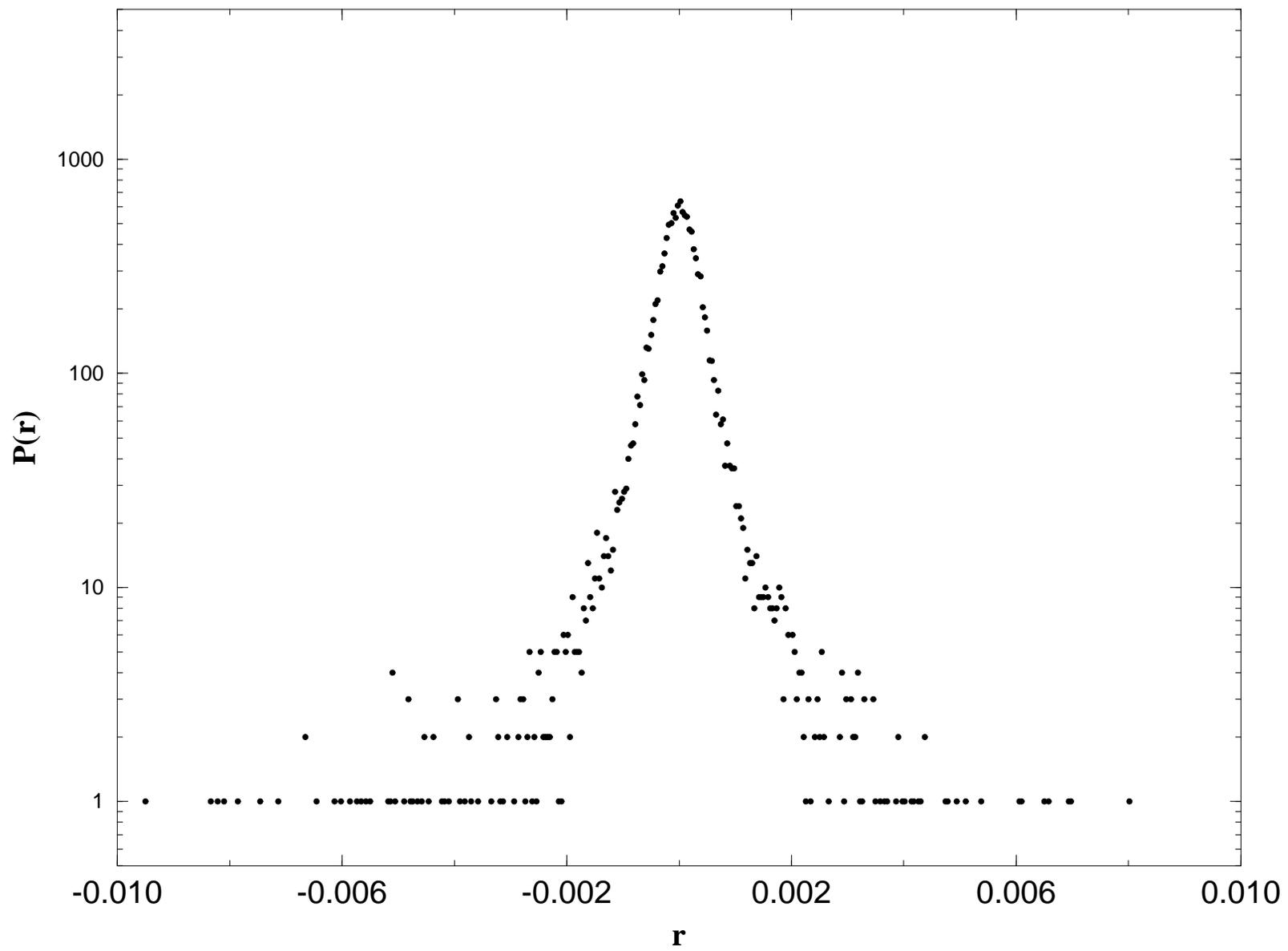
\begin{figure}
\caption{
The distribution of the variations of $\bar w$ after $\tau$ steps
$r(\tau) = [{\bar w(t+\tau)} - {\bar w(t)}] /{\bar w(t)}$,
where $\tau=50$, for the same parameters as in Fig.\ref{fig:pareto}.
This distribution has a L\'evy-stable shape with
$\alpha =1.4$. One can see that the shape on a semi-logarithmic scale differs
from a parabola (Gaussian distribution) in that it has significantly larger
probabilities for large $w_i$ values.
}
\label{fig:levy}
\end{figure}

Fig.
\ref{fig:levy}
shows the
distribution
P$(r)$
of the stochastic fluctuations
$r(\tau)$,
for
$\tau = 50$,
which is given
by a
L\'evy-stable distribution
$L_{\alpha}(r)$.
We find indeed that all the values of $r$
are smaller than the relative
width of
$\lambda$ (0.1)
times the maximal value
of
$w_i/{(N \bar w)}$
from Fig.
\ref{fig:pareto}.
The cut-off in the distribution
of the temporal fluctuations
originates
therefore in the cut-off in
the Pareto power law in
Fig.
\ref{fig:pareto}.
In the absence of this cut-off the variance of the distribution
of fluctuations would be infinite.
The divergence of the variance modulo finite size effects is analogous to the
divergence of the susceptibility in ordinary
statistical mechanics systems at criticality.

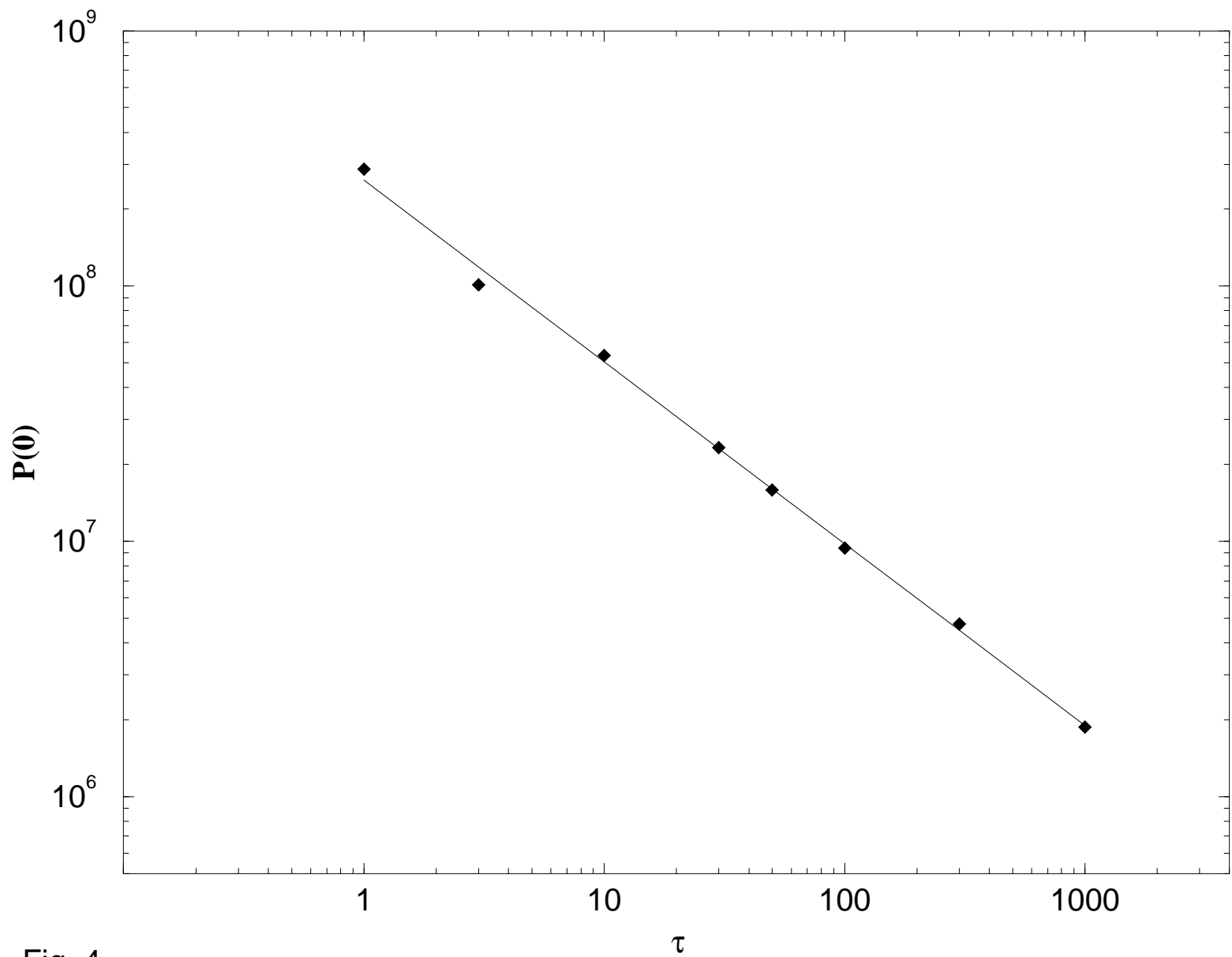
\begin{figure}
\caption{
The scaling with $\tau$ of the probability that
$r(\tau) = [{\bar w(t+\tau)} - {\bar w(t)}] /{\bar w(t)}$ is 0.
The parameters of the process are like in Fig. \ref{fig:pareto} and Fig. \ref{fig:levy}.
The slope of the straight line on the logarithmic scale is
$0.71$ which corresponds to a L\'evy-stable process with $\alpha = 1/0.71 = 1.4$.
}
\label{fig:levyzero}
\end{figure}

The peak
of the (truncated)
L\'evy-stable distribution
scales with $\tau$
according to

\begin{equation}
L_{\alpha}(r=0) \sim \tau^{-1/\alpha}
\label{fluct}
\end{equation}

\noindent
where $\alpha$ is the index of
the distribution.
In
Fig.
\ref{fig:levyzero}
we show the
height of the
peak
P$(r=0)$ of the distribution of fluctuations in
$\bar w$
as a function of
$\tau$
for the parameters used in Fig.
\ref{fig:pareto},
which give rise to a power-law distribution
of the $w_i$'s
with
$\alpha=1.4$.
It is found that the slope of the fit in Fig.
\ref{fig:levyzero}
is
$-0.71$
which is equal to
$-1/\alpha$,
following the scaling relation
of Eq.
(\ref{fluct}).
This is a further indication that the
fluctuations of
$\bar w$
follow a
L\'evy-stable distribution
with the index
$\alpha$
which equals the exponent of the Pareto power law in
Fig.
\ref{fig:pareto}.
It is gratifying that an explanation of the 100 years old Pareto power law in
these non-equilibrium systems which we have studied is provided by a straightforward
explicitation of the almost as old Lotka-Volterra
equation
\cite{Lotka25,Volterra26}.

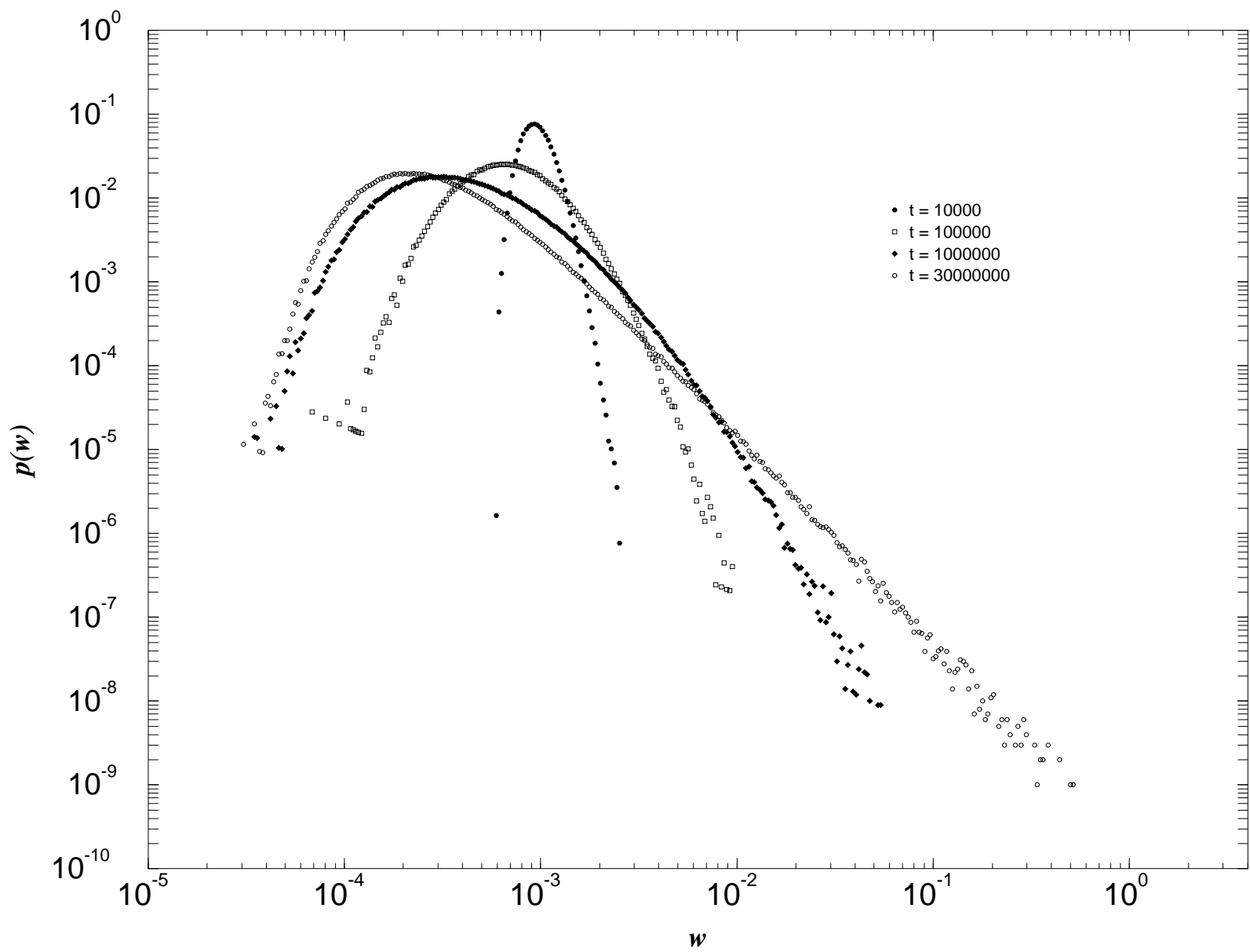
\begin{figure}
\caption{
The time evolution of $P(w)$ for a system starting from an uniform distribution of $w_i$.
In the first stages the distribution is log-normal and it then becomes power-law as
the non-multiplicative effects at the lower bound start being effective.
The process of convergence to the power law is much shorter than the actual equilibration
of the $\bar w$ value.
}
\label{fig:timeevolution}
\end{figure}

To provide more intuition
about the dynamics
leading to the power law distribution of $w_i$,
we show in
Fig.
(\ref{fig:timeevolution})
the time evolution of a GL system starting from a
uniform distribution of
$w_i$, $i=1,\dots,1000$.
We observe that the distribution gradually broadens.
In the first stages it becomes of log-normal form and then it
evolves into a
power-law as the non-multiplicative effects
in the vicinity of the lower bound become significant.

\section{Discussion}

\subsection{Previous Results}

The numerical results of the previous Section show convincingly that generic
(even non-stationary) systems with effective dynamics governed by the GL system of
Eq.(\ref{gedilo}), lead to (truncated) Pareto distributions of the system components.
They also lead to (truncated) L\'evy-stable laws of the fluctuations of the average. Let us
now explain intuitively why this is the case.
Consider first the Kesten system which is well known to present power
laws
\cite{Sornette97a,Sornette97,Sornett101,Sornett231,Sornett114,Kesten73}

\begin{equation}
w (t+1) = \lambda (t) w (t) + \rho (t)
\label{kesten}
\end{equation}
where the random numbers $\lambda$ and $\rho$ are extracted from two positive
distributions independent of $t$.
The Kesten system has a number of shortcomings which makes it unfit for most
practical applications in natural systems:
\begin{itemize}
\item In Eq. (\ref{kesten}), there is only one variable, (no index $i$). It
describes a non-interacting investor  (animal, city)
in a market (ecology, country) which induces effectively to him the return
(growth) $\lambda(t)-1$ after each trade (reproduction/replication/multiplication)
period $t$.

\item In order for this system to exhibit a power law distribution,
$\lambda$ has to be predominantly less than 1 such that it causes  $\lambda  w$
to be on average smaller than $w$.
Otherwise, the resulting $w$ distribution is a log-normal with width expanding
in time. This would correspond
in the infinite time limit
to a power law of the form $P(w) \sim w^{-1}$.
 The dependence of the Kesten model on a {\bf shrinking} dynamics is
incompatible with most of the natural systems in which the growth is positive.
For instance, the shrinking  multiplicative dynamics is certainly not a good model
for a stock market where the investors $i$ expect their wealth to increase on
average (otherwise they just stay out of the market).

\item

In realistic markets (ecologies,societies), the average wealth (population)
$\bar w$
varies significantly in time.
In the Kesten model this can be realized
only by varying the distributions of
$\lambda$
and
$\rho$
which in turn would
significantly affect the exponent
$\alpha$
of the power law
[Eq. (\ref{pareto})].
In the GL system, on the other hand, changes in the environment are
represented by changes in the coefficient $b$
of the resources limitation/competition term.
This can lead to changes by orders of magnitude in the total wealth/population
$N \bar w$ without affecting the exponent $\alpha$.
Interestingly, it turns out that the exponent
$\alpha$, in the distribution of wealth, has been stable for the last one hundred
years and across most western (capitalist) countries.
\end{itemize}

We will see later how the GL system solves the shortcomings of the Kesten model.
Meantime let us give an intuitive explanation of why the Kesten system leads to
a power law given by Eq. (\ref{pareto}).
First one should realize that due to the $\rho (t)$ term in Eq. (\ref{kesten}),
the values of $w(t)$ are typically kept above a certain minimal value of order
$\bar \rho$. Let us therefore effectively substitute the $\rho$ term in the Kesten
equation [Eq.(\ref{kesten})] with the condition that $w (t)  > {\bar \rho}$.
More precisely, each time $w (t)$ becomes smaller than $\bar \rho$, it is reposed
"by hand" to the value $\bar \rho$.

In the resulting system:
$w (t+1) = \lambda (t) w (t)$, with $w (t) > {\bar \rho}$
one can take the logarithm:
$ln w (t+1) = ln w (t) + ln \lambda (t)$.
The lower bound condition becomes then
$ln w (t) > ln {\bar \rho}$.
This represents a system in which $ln w$ undergoes a random walk with a drift
towards smaller values and with a reflecting barrier
at $ln {\bar \rho}$.
One can compare this with a molecule in gravitational field submitted to the
collisions with the rest of the gas (resulting in friction and Brownian motion)
and  bounded from below by the ground level.
It is not surprising therefore that (by analogy to the barometric equation)
the resulting probability distribution for $ln w$ is an exponential:

\begin{equation}
p(ln w) \sim e^{- \beta ln w}
\end{equation}

\noindent
which written in terms of $w$ itself gives:

\begin{equation}
P(w) \sim  w^{-1- \beta}
\end{equation}

\noindent
The particular value of $\beta$ depends in the Kesten system on the details of
the distribution
$\Pi (\lambda)$
and is such that the drift towards  lower values
induced by $\lambda$ is balanced by the drift to larger values induced by $\rho$.
As a consequence, this model, if (mistakenly) applied to the stock market
(ecology, society etc.), would predict not only {\bf negative} average returns
(growth) but also an exponent $\alpha$ in the power law that is highly sensitive
to the parameters
\cite{Sornette97a,Sornette97,Sornett101,Sornett231,Sornett114}.
On top of all these shortcomings, the Kesten system does not predict a (truncated) L\'evy-stable
distribution of the $\bar w$  fluctuations
(as repeatedly measured in nature
\cite{Cizeau143,Cizeau342,Mantegna94,Mantegna97,Mantegna96,Mantegna95,Stanley95,Stanley96,Amaral97a,Amaral97b}).
To get the (truncated) L\'evy-stable distribution
the following conditions should be satisfied:
the
index $i$, of the component
$w_i$ to be updated at time $t$,
is chosen randomly,
the
$w_i$'s satisfy a power-law distribution and the update
step is multiplicative, namely,
the change in $w_i$ is proportional to its current value.
For example, even if the dynamics leads to a power-law
distribution of $w_i$,
the fluctuation may not be described by L\'evy-stable distribution if
the magnitude of the update
of
$w_i$
is not proportional to
$w_i$ itself.

\subsection{How Does Our Model Work}

Let us now see how the GL model \cite{Solomon96a} solves the problems with the Kesten system.
The main new ingredients in the GL model are the appearance of $\bar w$ and
the appearance of an index $i$ in $w_i$. These two objects
allow the introduction of a
crucial ingredient which was absent in the Kesten system: the interaction
between the investors (sub-ecologies, sub-systems).
While the interaction in the stock market (ecology, society) is represented
in the Kesten system only implicitly by the stock returns (growth) $\lambda (t) - 1$
we introduce now additional interactions between the investors (individuals, families) $i$
which are mediated by the average $\bar w (t)$ and are crucial for the dynamics of the system.
Obviously, such terms, containing $\bar w (t)$ could not appear in an equation like Kesten
which considers only the dynamics of one variable at a time.
In order to introduce the crucial terms including $\bar w$ one has to give up the picture
of a single random investor and to embrace the picture of a macroscopic set of microscopic
investors, interacting among themselves through the market mechanisms.
The result is the system of nonlinear GL equations
which are
coupled through $\bar w(t)$
[Eq. (\ref{gedilo})].
In order to gain insight into the emergence of the power law and L\'evy-stable intermittency
in the GL system, one can express it formally as:
\begin{equation}
w_i (t+1) = \left[ \lambda (t) - b(t) {\bar w (t)} \right] w_i (t) + a(t)\bar w(t)
\label{tuning}
\end{equation}
If one ignores for the moment the effect of the changes of $w_i$'s on the
value of $\bar w$, the system [Eq. (\ref{tuning})] is of the Kesten type [Eq. (\ref{kesten})]
and we expect therefore the emergence of a scaling law, given by Eq. (\ref{pareto}).
If the effect of the changes in $w_i$ on $\bar w$ is considered, then one sees that
(for non-vanishing $b$) the system is self-tuning towards the value of $\bar w$
given by Eq. (\ref{equil}).
This self-tuning is realized by the dynamics of  the average in Eq. (\ref{tuning}).
If $\bar w (t)$ is small, then according the first term in Eq. (\ref{tuning}), $w_i$ will
typically increase and will make $\bar w (t)$ increase too.
If $\bar w (t)$ is large, then according the first term in
Eq. (\ref{tuning}), $w_i$ will typically decrease and will make $\bar w (t)$ decrease.
While in the synchronous Lotka-Volterra
[with a global time step updating based on Eq. (3)]
the system may have large steps and get into behavior alternating
chaotically between large and small $w(t)$ values,
in the case of the sequential
updating [Eq. (1)] of the $w_i$'s,
the average
will eventually self tune to a value of  $\bar w (t)$
given by Eq. (\ref{equil}).
The fluctuations around this value will be dominated by the first term in Eq.(\ref{tuning})
and will consist of a random walk with steps proportional to $w_i$.
Since $w_i$ are distributed by a power law, the fluctuations will be distributed by
a L\'evy-stable distribution of corresponding index
\cite{Levy37,Montroll82}.
In order to understand why the $w_i$ distribution is only weakly dependent on $b (t)$,
one can substitute Eq. (\ref{equil}) in to Eq. (\ref{tuning}) and use the normalized
variables $v_i = w_i/{\bar w}$.
One then obtains an equation of the Kesten form:
\begin{equation}
v_i (t+1) - v_i (t) \cong \left[ \lambda (t) - {\bar {\lambda} (t)} - a (t) \right] v_i (t) + a(t)
\label{tun}
\end{equation}
Note that we used here the approximation that the dynamics of $\bar w$
is much slower than the dynamics of $w_i$.
One observes that both $b (t)$ and $\bar w (t)$ are absent from Eq. (\ref{tun}):
their respective effects cancel.
In fact, one finds in simulations (Fig. \ref{fig:bzero}) that the distribution of $v_i (t)$ (and
therefore of $w_i(t)$) fulfills a power law of Eq. (\ref{pareto}) with exponent
$\alpha$ independent on the variations of $\bar w(t)$.
One also sees from Eq. (\ref{tun}) that the dynamics is invariant to an overall shift
in the distribution $\Pi (\lambda)$. This means that in particular the
GL multiplicative factors $\lambda$ can be significantly (and generically)
larger than unity allowing  (in contrast to the Kesten system) for expanding
(growing) dynamics.
Eq.
(\ref{tuning})
implies time correlations in the
{\it amplitude}
of the fluctuations
of
$\bar w$.
It was brought to our attention by D. Sornette that our data seem consistent with the
log-periodic corrections due to complex exponents discussed in \cite{sornette99}
as well as other the other data collected in the stock market
\cite{Cizeau143,Cizeau342,Mantegna94,Mantegna97,Mantegna96,Mantegna95,Stanley95,Stanley96,Amaral97a,Amaral97b,Bouchaud94,Matacz197,Potters98,Bouchaud97,cont97}.

Our mechanism relates the emergence of
power-laws and macroscopic fluctuations to the
existence of auto-catalyzing sub-sets in systems composed
of many microscopic entities.
In particular, the use of the $\bar w$ is not mandatory:
generic systems of the type
$w_i (t+1) = \sum_j \lambda_{ij} w_j (t)  - \sum_{j,k} b_{ijk} w_j (t) w_k (t)$
may also present similar properties.
At a more conceptual level,
the challenge is to identify,
in an as wide
as possible range of natural systems,
the elementary objects $i$,
the degrees of freedom $w_i$
associated with them
and the GL interactions explaining in each case the emergence of
scaling and intermittency.

\section{Summary}

In summary, we have studied
the dynamics of a
generic
class of
stochastic
Lotka-Volterra (discrete logistic) systems introduced in \cite{Solomon96a}
using computer simulations.
These systems consist of a large number of
interacting degrees of freedom
$w_i(t)$, $i=1,...N$,
which are
updated asynchronously.
The time evolution of each system component is dominated
by a stochastic individual autocatalytic dynamics,
in addition to a
global autocatalytic interaction mediated by the average
$\bar w(t)$, and a saturation term.
These models describe a large variety
of systems such as stock markets and city populations.
We find that the distribution $P(w,t)$
of  the system components $w_i$,
fulfills a Pareto power-law
$P(w,t)  \sim w^{-1-\alpha}$.
The average
${ \bar w (t)} $
exhibits intermittent fluctuations
following a (truncated) L\'evy-stable distribution
with the same index $\alpha$.
This intricate relation between the distribution of system
components and the temporal fluctuations resembles the
behavior of a variety of empirical systems.
For example, it provides a connection between the
power-law distribution of wealth in society
and the fluctuations in the stock market
which follow a (truncated) L\'evy-stable distribution.

\newpage

\noindent

\noindent

\noindent

\noindent

\noindent

\end{document}